\documentclass[floats,aps,prl,superscriptaddress,twocolumn,nofootinbib]{revtex4}
\usepackage{slashed}
\usepackage{amsmath}
\usepackage{graphicx}
\usepackage{epsfig}

\begin{document}

\title{A dynamical mechanism for generating quark confinement}

\author{Han-Xin He}
\affiliation{China Institute of Atomic Energy, P.O.Box 275(10), Beijing
102413, China}
\author{Yu-Xin Liu}
\affiliation{Department of Physics, Peking University,
Beijing 100871, China }
\begin{abstract}
We explore the dynamical mechanism for generating the infrared
singular quark-gluon vertex and quark confinement based on the
gauge invariance in covariant-gauge quantum chromodynamics(QCD).
We first derive the gauge-invariance constraint relation for the
infrared-limit behavior of the quark-gluon vertex, which shows the
mechanism for generating the infrared behavior of the quark-gluon
vertex. We hence unravel a novel mechanism for generating an
infrared singular quark-gluon vertex and then a linear rising
potential for confining massive quarks, where the infrared
singularity in the form factors composing the quark-ghost
scattering kernel plays a crucial role. The mechanism for linking
chiral symmetry breaking with quark confinement is also shown.


\end{abstract}
\maketitle

Quarks are the "elementary particles" of constituting hadrons,
which has been confirmed by experiments. However, the isolated
quarks have not been observed in nature. This phenomenon is called
quark confinement. To unravel the quark confinement from
underlying strong interaction theory quantum chromodynamics (QCD)
is doubtless one of major scientific challenges of this century.
Lattice QCD computations have derived a linear rising potential
between massive quarks, implying quark confinement at large
distances \cite{bali}, but the dynamical mechanism for generating
such a confining potential is still elusive. It is convinced that
the confinement mechanism should be encoded in the infrared(IR)
structure of QCD. This idea may be expressed by the notion of
infrared slavery\cite{wein}, i.e. the IR-singularities of QCD
Green functions generate confinement. Thus, to study the IR
behavior of QCD Green functions will be hopeful to unravel the
dynamical mechanism for generating confinement and dynamical
chiral symmetry breaking(DCSB), ect. In this aspect, the IR
behavior of the quark-gluon vertex and its relation to the
phenomena of quark confinement and DCSB are of particular
interest.

Intuitively, the best way to unravel the IR behavior of the
quark-gluon vertex would be to compute this vertex directly by
lattice QCD. However, at present this task seems to be hard to
perform since the complexity of the quark-gluon vertex\cite{skul}.
Here we suggest a practicable approach to explore the IR behavior
of the quark-gluon vertex based on the gauge invariance of QCD,
which is first to derive the gauge-invariance constraint relation
(like a sum rule) for the IR behavior of the quark-gluon vertex
relating to all of possible contributions to this IR-behavior from
the quark, gluon, ghost, and their interaction sectors. This is
similar to derive the gauge-invariant decomposition of the nucleon
spin into quark and gluon contributions for understanding the spin
structure of the nucleon\cite{jixi}. This approach allows one to
find the origins for generating the IR behavior of the quark-gluon
vertex and then to calculate their contributions respectively, and
hence to understand the dynamical mechanism for generating an
IR-singular quark-gluon vertex and quark confinement.

In this Letter, we explore the dynamical mechanism for generating
the IR behavior of the quark-gluon vertex and quark confinement in
covariant-gauge QCD according to above-suggested approach. We
first use the gauge-invariance of QCD, performed by using the
Slavnov-Taylor identity, to derive the constraint relation for the
IR-limit behavior of the quark-gluon vertex. This constraint
relation shows the mechanism that the IR behavior of the
quark-gluon vertex originates from two parts: the ghost
renormalization (the Yang-Mills sector) and the form factors
composing the quark-ghost scattering kernel (the quark-ghost
interaction sector). The gauge-invariance constraint relation for
the IR-limit behavior of the quark-gluon running coupling is also
derived. We then use these constraint relations to unravel the
dynamical mechanism for generating an IR-singular quark-gluon
vertex and a quark confining potential. The mechanism for linking
DCSB with quark confinement is also shown.

Let us first derive the gauge-invariance constraint relation for
the IR-limit behavior of the quark-gluon vertex. As is well-known,
the gauge invariance of QCD imposes powerful constraints on the
quark-gluon vertex, which are described by the generalized
Ward-Takahashi identity \cite{ward} or named the Slavnov-Taylor
identity (STI) \cite{slav}: The longitudinal part of the vertex
must satisfy the normal STI \cite{marc}, while the transverse part
of the vertex is constrained by the transverse STI \cite{hanx}. In
the limit of vanishing gluon momentum, the transverse part of the
vertex and then the transverse STI has no contribution. Thus, to
derive the constraint on the IR-limit behavior of the quark-gluon
vertex we only need the normal STI for the vertex which reads
\cite{marc}
\begin{eqnarray}
q^\mu \Gamma _{\mu}(k,p,q)G^{-1}(q^2)&=&
[S^{-1}(k)H(k,p,q)\nonumber \\
& & - \bar{H}(p,k,q)S^{-1}(p)],
\end{eqnarray}
where $q=k-p$, $q$ denotes the gluon momentum, $k$ and $p$ are
quark momentums, $S^{-1}(p)$ is inverse of the full quark
propagator, $G(q^2)$ is the ghost renormalization function
relating to the ghost propagator by $\tilde{D}(q)=-iG(q^2)/q^2$.
$H(k,p,q)$ is the quark-ghost scattering kernel and
$\bar{H}(p,k,q)$ is its "conjugate". The vertex $\Gamma_{\mu}$ is
related to the full quark-gluon vertex
$\Gamma_{\mu}^a(k,p,q)=gT^a\Gamma_{\mu}(k,p,q)$, where $T^a$ is
the generator of the gauge group $SU(N)$ with $N$ being the number
of colors, and $N=3$ for QCD. The matrix notation in both color
and Dirac matrices is understood.

The inverse of the full quark propagator in the Minkowski space
has the form
\begin{equation}
S^{-1}(p)= \alpha(p^2)\slashed{p} + \beta(p^2)I = Z_f^{-1}(p^2)(
\slashed{p}- M(p^2)),
\end{equation}
where $I$ is the unit matrix, $\slashed{p}=p^{\mu}\gamma_{\mu}$,
$\alpha(p^2)=Z^{-1}_f(p^2)$ and $\beta(p^2)=-M(p^2)Z_f^{-1}(p^2)$,
$M(p^2)$ is the quark mass function and $Z_f(p^2)$ denotes the
wave function renormalization. The DCSB is signaled by $M(p^2)\neq
0$ if the current quark mass $m=0$\cite{bowm}. The quark-ghost
scattering kernel $H(k,p,q)$ and its "conjugate" can be decomposed
in terms of scalar functions ("form factors")\cite{davy} as
\begin{eqnarray}
& &H(k,p,q)= \chi_0I + \chi_1 \slashed{k} + \chi_2 \slashed{p} +
\chi_3\sigma_{\mu
\nu}k^{\mu}p^{\nu},\nonumber \\
& &\bar{H}(p,k,q)=\bar{\chi}_0I + \bar{\chi}_1 \slashed{p} +
\bar{\chi}_2 \slashed{k} + \bar{\chi}_3\sigma_{\mu
\nu}k^{\mu}p^{\nu},
\end{eqnarray}
where $\sigma_{\mu\nu}=\frac{1}{2}[\gamma_{\mu},\gamma_{\nu}]$,
$\chi_i=\chi_i(k^2,p^2,q^2)$ and
$\bar{\chi}_i=\chi_i(p^2,k^2,q^2)(i=0,1,2,3)$. At the lowest
order, $\chi^{(0)}_0=1$ and $\chi_i^{(0)}=0$($i\geq1$). The
one-loop results for $\chi_i$ functions have been given already by
Ref.\cite{davy}.

To study how the STI imposes the constraint on the IR-limit
behavior of the quark-gluon vertex, we consider the limit of
vanishing gluon-momentum for the STI (1). At first, the right-hand
side of Eq.(1) can be written in terms of the decomposition forms
(2)-(3), which then can be calculated one-to-one after assuming in
the limit of vanishing gluon momentum $\chi_i$ and $\bar{\chi}_i$
behaving as $\lim_{q \rightarrow 0}\bar{\chi}_i=\lim_{q
\rightarrow 0}\chi_i \sim (q^2)^{\alpha_{\chi_i}}f_{(i)}(p^2)$. We
obtain
\begin{eqnarray}
& &{\lim_{q\rightarrow 0}}[S^{-1}(k)H(k,p,q)-\bar{H}(p,k,q)S^{-1}(p)]\nonumber \\
& &\sim {\sum_{i=0}^{3}}(q^2)^{\alpha_{\chi_i}+1/2}\hat{q}^{\mu}
{\sum_{j=1}^{3}}f_{(i,j)}(p^2)\tilde{L}_{\mu}^{(j)}(p),
\end{eqnarray}
where $\tilde{L}_{\mu}^{(1)}$=$\gamma_{\mu}$,
$\tilde{L}_{\mu}^{(2)}$=$p_{\mu}\slashed{p}$,
$\tilde{L}_{\mu}^{(3)}$=$p_{\mu}$, denoting the undressed tensor
structures for the longitudinal part of the quark-gluon vertex
(see following Eq.(9)), $\hat{q}^{\mu}=q^{\mu}/\sqrt{q^2}$ and
$f_{(0,1)}(p^2)=\alpha(p^2)f_{(0)}(p^2)$,
$f_{(1,1)}(p^2)=\beta(p^2)f_{(1)}(p^2)$, etc. Eq.(4) shows that
the right-hand side of this equation does not disappear if
$\alpha_{\chi_i}\leq -1/2$, i.e. $\chi_i$ ($\bar{\chi}_i$) being
IR-singular. But it does not mean that all $\chi_i$ and
$\bar{\chi}_i$ would be IR-singular. In fact, recent calculations
show that $\chi_0(\bar{\chi_0})$ is IR-finite \cite{agu1}. So far,
there is no calculation of $\chi_i(\bar{\chi_i})$ ($i\geq 1$).
From the structure of $H(\bar{H})$ given by Eq.(3), one may
conjecture that $\chi_1(\bar{\chi}_1)$ and $\chi_2(\bar{\chi}_2)$
may have similar IR-behavior, i.e,
$\alpha_{\chi_1}=\alpha_{\chi_2}$. To simplify discussion, let us
assume $\alpha_{\chi_1}=\alpha_{\chi_2}=\alpha_{\chi_3}$, thus
\begin{eqnarray}
& &{\lim_{q\rightarrow 0}}[S^{-1}(k)H(k,p,q)-\bar{H}(p,k,q)S^{-1}(p)]\nonumber \\
& &\sim (q^2)^{\alpha_{\chi_1}+1/2}\hat{q}^{\mu}{\sum_{i,j=1}^{3}}
f_{(i,j)}(p^2)\tilde{L}_{\mu}^{(j)}(p).
\end{eqnarray}
We note that the right-hand side of Eq.(5) will reduce to
$(q^2)^{\alpha_{\chi_i}+1/2}\hat{q}^{\mu}\sum_jf_{(i,j)}(p^2)\tilde{L}_{\mu}^{(j)}(p)$
with $i=1$ or 2 if only $\chi_1(\bar{\chi_1})$ or
$\chi_2(\bar{\chi_2})$ may have an IR-singularity.

The left-hand side of the STI (1) in the limit of vanishing gluon
momentum is easy to write after assuming a ghost renormalization
function behaving as $\lim_{q \rightarrow 0}G(q^2)\sim
(q^2)^{\alpha_G}$ and the quark-gluon vertex possessing the IR
power exponent $\delta_{qg}$. We have
\begin{eqnarray}
& &{\lim_{q\rightarrow 0}}q^{\mu}\Gamma _{\mu}(k,p,q)G^{-1}(q^2)\nonumber \\
& &\sim
(q^2)^{\delta_{qg}+1/2-\alpha_G}\hat{q}^{\mu}{\sum_{j=1}^{3}}f_{(j)}(p^2)
\tilde{L}_{\mu}^{(j)}(p).
\end{eqnarray}

Equating Eq.(5) with Eq.(6) for a definite momentum $p$ and
comparing their power exponent, we thus find the constraint
relation for the IR-limit behavior of the quark-gluon vertex,
imposed by the gauge invariance, as
\begin{eqnarray}
& &\lim_{q \rightarrow 0}\Gamma_{\mu}(k,p,q) \sim
(q^2)^{\delta_{qg}}\tilde{L}_{\mu}(p)\nonumber \\
& & \quad {\rm with~} \delta_{qg}=\alpha_{\chi_i}+\alpha_G,
\end{eqnarray}
where $i\geq1$,
$\tilde{L}_{\mu}(p)=\sum_{j=1}^{3}f_{(j)}(p^2)\tilde{L}_{\mu}^{(j)}(p)$
and $f_{(j)}(p^2)=\sum_if_{(i,j)}(p^2)$. The constraint relation
(7) shows a "sum rule" which indicates the mechanism that the
IR-behavior of the quark-gluon vertex is generated by two parts:
the form factors composing the quark-ghost scattering kernel and
the ghost renormalization function. This constraint relation is
the central result of the IR analysis for the quark-gluon vertex.

The further interesting quantity is the nonperturbative running
coupling from the quark-gluon vertex, which is defined by the
nonperturbative and renormalization group invariant relation
\begin{equation}
\alpha_{qg}(\mu^2)=\frac{g^2_0}{4\pi}Z_{1F}^{-2}(\mu^2)Z^2_f(\mu^2)Z(\mu^2).
\end{equation}
Here $Z(\mu^2)$ is the gluon renormalization defined in the
renormalization scale $\mu$, $Z_{1F}(\mu^2)$ is the
renormalization function of the quark-gluon vertex defined
according to
$\Gamma_{\mu,R}=Z_{1F}\Gamma_{\mu}=Z_{1F}\lambda_1\gamma_{\mu}+
...$, where $\lambda_1$ dresses the $\gamma_{\mu}$-part of the
vertex, which gives $Z_{1F}(\mu^2)=\lambda_1^{-1}(\mu^2)$.

To write $\lambda_1(\mu^2)$, we use the Ward-type identity for the
quark-gluon vertex\cite{han2}, which can be derived based on the
STI (1) in the limit of vanishing gluon momentum. The result shows
that the quark-gluon vertex in the limit of vanishing gluon
momentum reads
\begin{eqnarray}
\Gamma_{\mu}(p,p) &=& \lambda_1(p^2)\gamma_{\mu} +
\lambda_2(p^2) \slashed{p}p_{\mu}\nonumber \\
& & + \lambda_3(p^2)p_{\mu} + \lambda_4(p^2)
\slashed{p}\gamma_{\mu},
\end{eqnarray}
where $\lambda_i(p^2)\equiv \lim_{q \rightarrow 0}
\lambda_i(k^2,p^2,q^2)(i=1,2,3)$($\lambda_4(p^2)=0$). After
setting $p=\mu=q$ we have
\begin{equation}
\lambda_1(q^2)=Z_f^{-1}(q^2)F_{1,H}(q^2)G(0),
\end{equation}
where $G(0)=\lim_{q \rightarrow 0}G(q^2)$ and
\begin{eqnarray}
F_{1,H}(q^2)
&=&\chi_0(q^2)-2q^2\chi_3(q^2)\nonumber \\
& &+M(q^2)[\chi_2(q^2)-\chi_1(q^2)].
\end{eqnarray}
We thus find
\begin{eqnarray}
\alpha_{qg}(q^2)&=&\frac{g_0^2}{4\pi}F^2_{1,H}(q^2)G^2(0)Z(q^2)\nonumber \\
&=&F^2_{1,H}(q^2)(G(0)/G(q^2))^2\alpha_{gh}(q^2),
\end{eqnarray}
where $\alpha_{gh}(q^2)=\frac{g_0^2}{4\pi}G^2(q^2)Z(q^2)$, which
is the ghost-gluon running coupling \cite{bouc} .

Eq.(12) indicates that the quark-gluon running coupling is related
to the ghost-gluon running coupling through the vertex structure
factor $F_{1,H}(q^2)$. The IR-limit behavior of $F_{1,H}(q^2)$ can
be obtained by using Eq.(11). After assuming
 $\lim_{q \rightarrow
0}F_{1,H}(q^2)\sim (q^2)^{\alpha_H}$ and $\lim_{q \rightarrow
0}M(q^2)\sim (q^2)^{\alpha_M}$, we then have
$\alpha_H=\alpha_{\chi_i}+\alpha_M$, where we used the result that
 $\chi_0$ is IR-finite\cite{agu1}. Thus, by Eq.(12) we further
find the gauge-invariance constraint relation for the IR behavior
of the quark-gluon running coupling in the limit of vanishing
renormalization scale:
\begin{eqnarray}
& &\lim_{q \rightarrow 0}\alpha_{qg}(q^2)  \sim
(q^2)^{\delta_{\alpha_{qg}}} \nonumber \\
& &\quad {\rm with~} \delta_{\alpha_{qg}}=
2\alpha_{\chi_i}+2\alpha_M+\delta_{\alpha_{gh}},
\end{eqnarray}
where $\delta_{\alpha_{gh}}=2\alpha_G+\alpha_Z$ denoting the
IR-power exponent of the ghost-gluon running coupling. The
constraint relation (13) indicates the mechanism for generating
the IR-behavior of the quark-gluon running coupling.

Now we study how the constraint relations (7) and (13) relate to
the dynamical mechanism for generating an IR-singular quark-gluon
vertex and quark confinement. To answer this question, we need the
knowledge about $\alpha_G$, $\delta_{\alpha_{gh}}$ $\alpha_M$ and
$\alpha_{x_i}$. Note that constraints (7) and (13) can be used for
any covariant gauge. So far, the IR behavior of the Yang-Mills
Green functions has been intensely studied in Landau gauge and two
types of solutions have been found: The scaling solution gives
$\alpha_G=-\kappa\simeq -0.595$ and $\delta_{\alpha_{gh}}=0$,
while the massive gluon (decoupling) solutions show $\alpha_G=0$
and $\delta_{\alpha_{gh}}\simeq 0$ if the massive gluon behavior
is taken into account ( Normally $\delta_{\alpha_{gh}}=1$)
\cite{bouc}. One may consider the scaling solution as "critical"
one with a $unique$ (critical) value of the coupling for which
$G(0)\rightarrow \infty$, while the decoupling solutions as a
family of sub-critical ones with finite $G(0)$\cite{bouc}. Thus,
although recent computations seem to favor the decoupling-type
solutions, the scaling solution is still valuable for our
discussions. The quark mass function $M(q^2)$ is
IR-finite\cite{bowm} and so $\alpha_M=0$. As for $\alpha_{x_i}$,
there is no calculation result for $\alpha_{x_i}(i\geq 1)$. We
shall separately discuss its role for the case
$\alpha_{x_i}=-1/2$, $\alpha_{x_i}<-1/2$ and $\alpha_{x_i}>-1/2$.
Based on above information, we now apply the constraint relations
(7) and (13) to derive the interaction potential between quarks in
Landau gauge. We first consider the case that the form factor
$\chi_i$ has the IR-singularity with $\alpha_{\chi_i}=-1/2$ and
DCSB appears ($M\neq 0$). In this case we have
\begin{eqnarray}
& &\lim_{q \rightarrow 0}\Gamma_{\mu}(k,p,q) \sim
(q^2)^{-1/2+\alpha_G}\tilde{L}_{\mu}(p),\nonumber \\
& &\lim_{q \rightarrow 0}\alpha_{qg}(q^2) \sim
(q^2)^{-1+\delta_{\alpha_{gh}}},
\end{eqnarray}
which generate an IR singular quark-gluon vertex and an IR
divergent quark-gluon running coupling in the scaling case. Such
an IR-singular quark-gluon vertex or quark-gluon running coupling
with nonperturbative one-gluon exchange between two massive quarks
generate an interaction quark potential
\begin{equation}
\lim_{q \rightarrow 0}V(q) \\ \sim \\ \frac{1}{(q^2)^2}
\end{equation}
in the scaling case. This quark potential in the coordinate space
is normally written
\begin{equation}
V(\vec{r})=\int
\frac{d^3q}{(2\pi)^3}V(q^0=0,\vec{q})e^{i\vec{q}\cdot\vec{r}}
\\\sim \\\mid \vec{r} \mid,
\end{equation}
generating a linearly rising potential between massive quarks,
implying quark confinement at large distances.

The constraint relations (7) and (13) are also valid in the chiral
symmetry phase of QCD. In this case $M=0$, the factor
$F_{1,H}(q^2)$ becomes IR-finite due to the disappearance of the
term $M(q^2)[\chi_2(q^2)-\chi_1(q^2)]$ with IR-singularity (i.e,
$f_{(1,1)}(p^2)=0, f_{(2,1)}(p^2)=0$) and hence $\alpha_H=0$. The
constraint relations (7) and (13) then reduce to
$\delta_{qg}=\alpha_G$ and
${\delta_{\alpha_{qg}}}={\delta_{\alpha_{gh}}}$, which imply that
the quark-gluon vertex has similar IR-limit behavior as that of
the ghost renormalization function and the quark-gluon running
coupling in IR-limit goes to a fixed point similar to the
ghost-gluon running coupling. Such king of solution no longer
leads to a confining potential.

Thus we have unravelled a novel mechanism for generating an IR
singular quark-gluon vertex and quark confinement as well as a
mechanism for linking DCSB with quark confinement and chiral
symmetry restoring with the disappearance of confinement.

It is interesting to notice that the IR singular behavior of the
quark-gluon vertex and the quark-gluon running coupling given by
Eq.(14) in the scaling case is consistent with the result given by
Ref.\cite{alko}. However, the mechanism for generating such kind
of IR singular behavior is different. Present work shows that by
gauge-invariance constraints the IR singularity scaling with
$(q^2)^{-1/2}$ of the form factors $\chi_i$($i\geq1$) composing
the quark-ghost scattering kernel $H$ plays a crucial role in
generating such kind of IR singular quark-gluon vertex,
especially, the quark-gluon running coupling scaling with
$(q^2)^{-1}$. Accordingly, the $H$ would be IR-singular and the
singularity with $\alpha_{\chi_i}=-1/2$ in $H$ leads to that the
right-hand side of STI (1) is just an IR constant (see Eq.(5) for
a definite $p$) but not IR-singular. This is in sharp contrast to
the power counting scheme of Ref.\cite{alko} where such kind of IR
singular behavior is an induced effect from singularities in the
pure Yang-Mills theory and $H$ is IR-finite.

We note that in the decoupling cases using Eq.(14) can not lead to
a linear confining potential at large distances. Does it mean that
the mechanism to explain confinement would be different for the
decoupling cases? This needs to be studied further. We should
notice that the different IR behavior of the ghost (gluon)
propagator for the scaling case and decoupling cases sets in at
scales $q^2 \ll \Lambda^2_{QCD}$, while the difference between a
scaling and a massive gluon (decoupling) solution is far less
marked at scales $q^2 \geq \Lambda^2_{QCD}$ and all dynamics
describing the physics of hadron world takes part on such
scales\cite{bouc,penn}. Hence it is reasonable to think that a
linear confining potential also holds in decoupling cases, like
the scaling case, at scales $q^2 \geq \Lambda^2_{QCD}$. We thus
can infer that while a linear confining potential is generated at
scales $q^2 \geq 0$ by Eq.(14) with nonperturbative one-gluon
exchange in the scaling case with a "critical" value
$G(0)\rightarrow \infty$, a linear confining potential may be
generated also at scales $q^2 \geq \Lambda^2_{QCD}$ by the same
way in decoupling cases with the sub-critical finite $G(0)$, where
the IR-singularity in the quark-gluon vertex, generated by the
IR-singularity of the form factors $\chi_i(i\geq1)$ with
$\alpha_{\chi_i}=-1/2$ , plays a crucial role for the generation
of the confining potential.

Now let us return to discuss the possible IR-behavior of the
quark-gluon vertex in the case $\alpha_{\chi_i}>-1/2$ or
$\alpha_{\chi_i}<-1/2$. The case $\alpha_{\chi_i}<-1/2$ gives the
constraint $\delta_{qg}<-1/2+\alpha_G$. In such case, an
IR-singular quark-gluon vertex scaling as $(q^2)^{-1+\alpha_G}$
might exist if $\alpha_{\chi_i}=-1$, which can lead to a linear
confining potential at large distances in the decoupling cases,
but the corresponding quark potential is given by $V(\vec{r}) \sim
|\vec{r}|^3$ at large distances in the scaling case. Such a case
looks not reasonable. Interestingly, the case
$\alpha_{\chi_i}>-1/2$ gives the constraint
$\delta_{qg}>-1/2+\alpha_G$, where an IR-regular quark-gluon
vertex may appear if all functions $\chi_i$ are IR-finite.

Finally, we would like to point out that Eq.(12) provides an
approach to calculate the quark-gluon running coupling
perturbatively and nonperturbatively. Thus it will be able to
build a united description for the scale evolution of the
quark-gluon running coupling from the ultraviolet momentum region
up to the IR-momentum region, helping us to intuitively understand
how the QCD interaction strength changes from the asymptotic
freedom region to nonperturbative region.

To summarize, we explored the dynamical mechanism for generating
the IR-singular quark-gluon vertex and quark confinement based on
the gauge invariance in covariant gauge QCD. We have first derived
the gauge-invariance constraint relations for the IR behavior of
the quark-gluon vertex and the quark-gluon running coupling by
using the STI for the vertex, which gives the mechanism for
generating the IR behavior of the quark-gluon vertex and the
quark-gluon running coupling. We hence unravelled a novel
mechanism for generating an IR singular quark-gluon vertex and
then a linear confining potential as well as a mechanism for
linking DCSB with quark confinement and chiral symmetry restoring
with the disappearance of confinement, where the IR-singularity
with $\alpha_{\chi_i}=-1/2$ of form factors $\chi_i(i\geq1)$
composing the quark-ghost scattering kernel plays a crucial role.
Above picture can be tested, which calls for consistent
calculations of form factors. But at present one can not yet
exclude the possibility that the quark-gluon vertex may be
IR-regular provided all functions $\chi_i$ are IR-finite.
Therefore, to study the IR behavior of the form factors composing
the quark-ghost scattering kernel, by consistently calculating
those form factors, then becomes a crucial test to unravel the
real IR behavior of the quark-gluon vertex and the picture of
quark confinement dynamics.

We are grateful to C.D.Roberts for valuable comments. We also
thank X.D.Ji, H.S.Zong and H.L.Ma for useful discussions. This
work is supported by the National Natural Science Foundation of
China under grant Nos.10935001, 11075052, 11175004, and 11175262.

\end{document}